\documentclass[11pt,twoside]{article}
\usepackage{asp2004}
\usepackage{psfig}
\usepackage{epsf}
\usepackage{graphics}
\usepackage{lscape}
\usepackage{xspace}
\def\edcomment#1{\iffalse\marginpar{\raggedright\sl#1\/}\else\relax\fi}
\reversemarginpar

\newcommand{\ASCA}{{\it ASCA}\xspace}

\newcommand{\XMM}{{\it XMM-Newton}\xspace}

\newcommand{\CHANDRA}{{\it Chandra}\xspace}

\newcommand{\UNITLUMI}{{\rm ergs~s$^{-1}$}\xspace}

\newcommand{\UNITNH}{{\rm cm$^{-2}$}\xspace}
\newcommand{\UNITVEL}{{\rm km~s$^{-1}$}\xspace}

\newcommand{\ARCSEC}{{$''$}\xspace}

\newcommand{\NH}{{\it N$_{\rm H}$}\xspace}

\newcommand{\LX}{{\it L$_{\rm X}$}\xspace}

\newcommand{\KT}{{\it kT}\xspace}

\newcommand{\EM}{{\it E.M.}\xspace}

\begin{document}
\title{\XMM Observations of the 2003 X-ray Minimum of $\eta$ Carinae}
\author{K. Hamaguchi$^{1,2}$, M.~F. Corcoran$^{1,3}$, T. Gull$^{4}$, N.~E. White$^{1}$, A. Damineli$^{5}$, K. Davidson$^{6}$}
\affil{1: NASA/GSFC/LHEA, Greenbelt, MD 20771,
2: National Research Council, 500 Fifth Street, NW, Washington, D.C. 20001,
3: Universities Space Research Association, 7501 Forbes 
Blvd, Ste 206, Seabrook, MD 20706,
4: NASA/GSFC/LASP, Greenbelt, MD 20771,
5: Instituto Astron{\^ o}mico e Geof{\' i}sico da USP, 
R. do Matao 1226, 05508-900, S{\~ a}o Paulo, Brazil,
6: Astronomy Department, University of Minnesota, 116 Church Street SE, Minneapolis,
MN 55455
}

\begin{abstract}
The \XMM X-ray observatory took part in the 
multi-wavelength observing campaign of the massive, evolved 
star $\eta$ Carinae in 2003 during its recent X-ray minimum.
This paper reports on the results of these 
observations, mainly from the aspect of spectral change.
Hard X-ray emission from the point source of $\eta$ Carinae was 
detected even during the minimum. During the minimum the observed flux above 3 
keV was $\sim$3$\times$10$^{-12}$ ergs cm$^{-2}$ s$^{-1}$, which is 
about one percent of the flux before the minimum. 
Changes in the spectral shape revealed two X-ray emission components in the central
point source.
One component is non-variable and has relatively cool plasma of $kT\sim$1~keV
and moderate absorption, $N_{H}\sim$5$\times$10$^{22}$~\UNITNH.
The plasma is probably located far from the star, possibly produced by
the high speed polar wind from $\eta$ Carinae.
The other high temperature component has $kT\sim5$~keV and is strongly variable.  This component shows 
an increase in the apparent column 
density from 5$\times$10$^{22}$~\UNITNH to 2$\times$10$^{23}$~\UNITNH,
probably originating near the heart of the binary system.
These changes in \NH were smaller than expected if the minimum is produced solely by an
increase of hydrogen column density.
The X-ray minimum seems to be dominated by a decrease of the apparent
emission measure, suggesting that the brightest part of the 
X-ray emitting region is completely obscured during the 
minimum in the form of an eclipse. A ``partial covering'' model might explain the residual emission seen during the minimum.
\end{abstract}
\thispagestyle{plain}
\section{Introduction}
The supermassive star, $\eta$ Carinae, is now widely described
as a binary system
with a period of 5.5~years \citep[e.g.][]{Ishibashi1999,Damineli2000}.
A collision between the wind from the primary star and the hidden companion forms a strong bow shock,
which will produce hot X-ray emitting plasma.
The emission undergoes a flux minimum apparently coincident with 
periastron passage, which 
may be caused by partial or full eclipse by the wind of the primary star and/or
collapse of the bow shock.
During the minimum in 1997$-$98,
the \ASCA satellite detected hard X-ray emission, characterized by similar \KT and \NH to
the pre-minimum state 
with reduced
plasma emission measure (\EM).
The limited spatial resolution of \ASCA, however, left the possibility that 
the observed emission 
was contaminated by emission from 
unresolved nearby sources.

The latest X-ray minimum began at 2003 June 29.
It was monitored with three X-ray observatories, RXTE, \CHANDRA, and \XMM
(see Corcoran et al., this meeting, for the results of the RXTE and \CHANDRA\ campaigns.)
Of these observatories, \XMM has the largest effective area, with moderate spatial and spectral resolution,
and therefore is suitable for tracing  changes in \NH and \KT.
\XMM observed $\eta$ Carinae 
1) 5 times in January 2003 before the minimum, 
which are treated as one set of data in our analysis, 
2) twice in June 2003 prior to the minimum, near the X-ray maximum, 
3) four times during the minimum in July and August 2003, for a total of 11 observations.
This paper reports on the results 
from the EPIC  pn and MOS CCD detectors.

\begin{figure}
\plotone{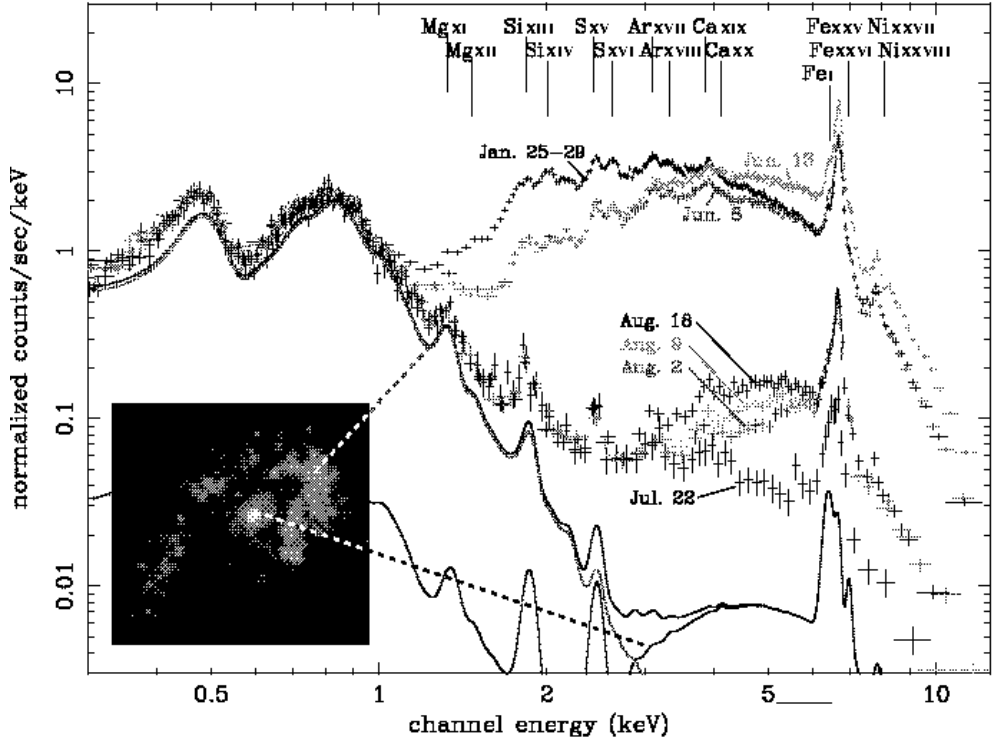}
\caption{EPIC pn spectra of $\eta$ Carinae in 2003.
Spectra from each observation are shown as individual points with error bars.
Spectra of the ``Outer Debris Field'' and the Homunculus Nebula, and their summed spectrum
are shown as solid lines.
The bottom left panel displays a \CHANDRA image during the X-ray minimum \citep{Corcoran2004}.
Strong emission lines from individual elements are shown at the top of the figure.}
\label{fig:xmmspectra}
\end{figure}

\section{Results}
A \CHANDRA observation by \citet{Corcoran2004} confirmed that hard X-ray 
emission during the X-ray minimum comes mainly from a point source at the position of
$\eta$ Carinae and partly from 
faint X-ray emission reflected from the Homunculus Nebula around $\eta$ Carinae
(see the bottom left panel in Figure~\ref{fig:xmmspectra}).
Unfortunately, \XMM cannot 
spatially
resolve these components nor well separate the 
soft X-ray emission from the ``Outer Debris Field'' beyond the Homunculus,
which is made by ancient ejecta interacting with the interstellar medium or 
previous ejecta.
We therefore included all these components in the source region when extracting source events,
and took background events from source free regions on the same CCD chip.

Background subtracted light curves above $\sim$2~keV
exhibited 
variation up to $\sim$5\% on timescales of $\sim$30 ksec,
which reflects emission from the central point source.
The variation 
roughly agrees with
the 
interpolated 
daily fluxes monitored with RXTE

\begin{figure}[t]
\plotone{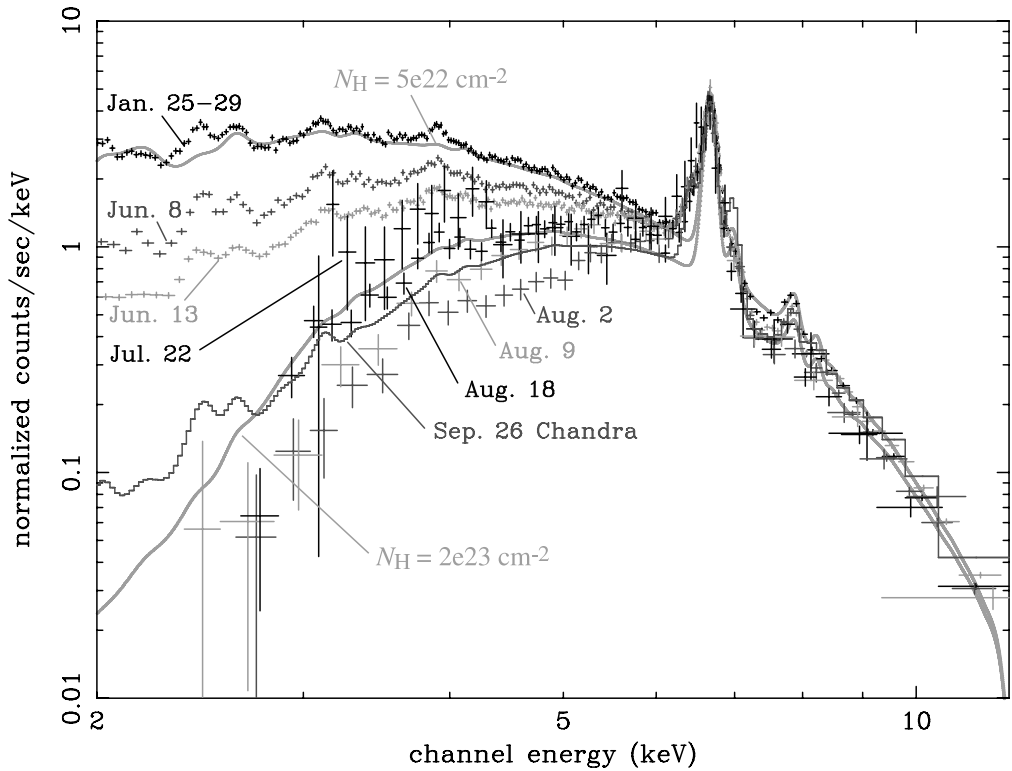}
\caption{\XMM EPIC pn spectra and the best-fit model of the \CHANDRA spectrum on September 26,
normalized at the Fe$_{\rm XXV}$ line intensity.
The vertical unit is correct for the data on January 25--29.
Smooth gray lines with indications of \NH
represent absorbed 1T thin-thermal plasma (MeKaL) models with \KT =5 keV,
\NH =5$\times$10$^{22}$\UNITNH (upper line) and =2$\times$10$^{23}$ \UNITNH (lower line),
\EM =1.7$\times$10$^{57}$ cm$^{-3}$ and abundance $Z/Z_{\odot}=0.87$.}
\label{fig:xmmnormspectra}
\end{figure}

The EPIC pn spectra in Figure~\ref{fig:xmmspectra} demonstrate 
a
strong decrease
in hard X-ray emission up to a factor of $\sim$100
above 1~keV during the minimum,
with
no variation in the emission below
1~keV.
This is 
similar to the \ASCA results 
from
the previous X-ray cycle
\citep{Corcoran2000}.
The contribution of the X-ray
emission from the outer debris field
and the Homunculus Nebula,
neither of which varied
on timescales of a month or longer,
was
estimated 
from
the \CHANDRA data on 2003 July 20 during the minimum,
and is overlaid in Figure~\ref{fig:xmmspectra}.
Except for the excess below 1~keV
(which is produced by the poor absolute flux calibration between \CHANDRA\ and \XMM),
the excess above 1~keV should 
represent
emission from the central source.

During the minimum, this emission shows
a non-variable component
between 1--3~keV.
To estimate the physical 
properties of this non-variable component,
we subtracted the 
X-ray emission from the outer debris field and the reflected emission from the Homunculus
from the spectra during the minimum, 
and simultaneously fit them
by a two-temperature (2T) optically thin-thermal plasma (APEC) model with
independently absorbed 1T non-variable and 1T variable components.
We also used the \CHANDRA spectra of the central region for the fitting.
The best-fit parameters of the non-variable component are \KT $\sim$1~keV, 
\NH $\sim$5$\times$10$^{22}$~\UNITNH, and log~\LX $\sim$34.2~\UNITLUMI.

We extracted spectra of the variable component by subtracting 
this
non-variable component in addition to the X-ray 
debris field and Homunculus emission components
(Figure~\ref{fig:xmmnormspectra}).
In the figure, we also display the best-fit model of the \CHANDRA spectrum from September 26 
when the minimum had just ended, and normalized all the spectra
at the Fe$_{\rm XXV}$ line energy.
Interestingly, the normalized spectra do not show any significant change in
hard band slope above 7~keV, which is equivalent to a continuum temperature of \KT $\sim$5~keV.
Meanwhile, the relative flux in the lower energy band decreased, 
with \NH increasing from 5$\times$10$^{22}$~\UNITNH 
to 2$\times$10$^{23}$~\UNITNH.
The \NH increase was moderate, and is not large enough to account
entirely for the strong flux decrease during the minimum,
especially at energies $>3$~keV.
The flux decrease is more consistently described as an apparent decrease of \EM 
as suggested in the earlier \ASCA observations \citep{Corcoran2000}.  
This could either represent a real reduction in the amount of X-ray emitting plasma, or an obscuration of X-ray emitting plasma.
The flux  during the minimum started to increase 
around July 22, though 
\NH was still increasing, and \NH continued to increase
through the recovery on September 26.
The \NH increase seems to lag
the apparent \EM decrease.
In the spectra near the X-ray intensity maximum,
the Fe$_{\rm XXV}$ emission line seems to have a lower energy tail, which is
also seen in the \CHANDRA HETG high resolution spectra 
(see Corcoran et al., this meeting).
The feature seemed to be enhanced during the minimum though photon statistics 
were rather limited.
This low energy tail may suggest that 
the ionization of the X-ray emitting plasma during the X-ray minimum may have been out of collisional equilibrium.
The EWs of the Fe fluorescent line were 140--220~eV before the minimum and were
restricted to less than 700~eV during the minimum.

\section{Discussion}

\subsection{What is the Non-Variable X-ray Source?}

The non-variable X-ray emission was stable for about two months during the 
minimum when the X-ray emission from the colliding winds exhibited prominent variation.
The \NH of $\sim$5$\times$10$^{22}$~\UNITNH is smaller than the columns to the colliding wind
X-rays around the minimum and are
the same as those in January and near apastron \citep[e.g.][]{Leutenegger2003}.
These results  
suggest that this emission component is remote from the binary system and not affected by the increasing column to the colliding wind source.
On the other hand, \CHANDRA images during the minimum between 1$-$3~keV,
dominated by the non-variable X-ray emission,
did not show any extended structure.
The plasma size is therefore restricted to within $\sim$1\ARCSEC, equivalent to 
a physical size of $\sim$2300~AU assuming $d \sim$2.3~kpc.
These results suggest that the X-ray plasma is produced by collision of a fast outflow
with ambient gas relatively far from the star.
A good candidate for this outflow is the polar wind from $\eta$ Carinae, which
has a high speed outflow up to $\sim$1000 \UNITVEL \citep{Smith2003},
which can produce plasma at temperatures near 1~keV.

\subsection{What Caused the X-ray Minimum?}

The series of the \XMM observations confirmed that 
the spectral variations during the minimum are caused by a
change in \EM, and not the observed increase in \NH (which is too low to provide the observed 
decline at $E>5$ keV).
There are a number of ways in which this apparent reduction in X-ray brightness can be produced.
One scenario is that the X-ray activity decayed close to the periastron passage, perhaps due to strong instabilities near periastron \citep{Davidson2002}.
However, while the behavior of the Fe$_{\rm XXV}$ line suggests that the ionization balance of the plasma may be  changing during the minimum,
the hottest plasma temperature appears so stable that it does not support
a dramatic decay of X-ray activity.
An alternative is that the X-ray emission is only partially blocked by an optically thick absorber.
In this scenario, the plasma \EM is apparently reduced during the X-ray minimum
because the volume is mostly obscured from our line of sight.
A partial or annular eclipse geometry seems unlikely  because
of the rapid change in geometry caused by the motion of the companion near periastron.
A possible solution is a "leaky absorber" consisting of optically thick clumps immersed in a 
lower-density gas  which obscures 95--99\% of the emitting region.



\end{document}